\documentclass[floatfix,twocolumn,aps,prl,showpacs]{revtex4-1}
\usepackage{graphicx}
\usepackage{amsmath}
\usepackage{amssymb}

\newcommand{\threevec}[3]{\left(\begin{array}{c}#1\\#2\\#3\end{array}\right)}
\newcommand{\bvec}[1]{\mathbf{#1}}
\newcommand{\nematic}{\bvec{\hat{d}}}
\newcommand{\absF}{|\langle\bvec{\hat{F}}\rangle|}
\newcommand{\inleva}[1]{\langle#1\rangle}
\newcommand{\abs}[1]{\left|#1\right|}
\newcommand{\eva}[1]{\left<#1\right>}
\def\SO{{\rm SO}}
\def\U{{\rm U}}

\begin{document}

\author{Magnus O.\ Borgh}
\email{M.O.Borgh@soton.ac.uk}
\author{Janne Ruostekoski}
\email{janne@soton.ac.uk}
\affiliation{School of Mathematics, University of Southampton, SO17 1BJ,
    Southampton, UK}

\title{Topological interface engineering and defect crossing in
  ultracold atomic gases}

\begin{abstract}
We propose an experimentally feasible scheme for topological interface
engineering and show how it can be used for studies of dynamics of
topologically nontrivial interfaces and perforation of defects and
textures across such interfaces. The method makes use of the internal
spin structure of the atoms together with locally applied control of
interaction strengths to create many-particle states with highly
complex topological properties. In particular, we consider a
constructed coherent interface between topologically distinct phases
of spinor Bose-Einstein condensates.
\end{abstract}
\pacs{%
03.75.Lm, 
03.75.Mn, 
67.85.Fg, 
11.27.+d, 
}
\maketitle

At the interface of two topologically distinct phases of a
macroscopically coherent quantum system, the
symmetry properties of the ground-state wave function
change. Topological defects (e.g., vortices) cannot in general
penetrate the interface
unchanged. The boundary therefore has the property that
defects must either terminate on the interface (typically as a point defect or
monopole)
or connect nontrivially to another object on the opposite side of the
boundary.

Interfaces between topologically distinct regions play an important
role, e.g., in exotic
super\-conductivity~\cite{bert_nphys_2011},
superfluid liquid $^3$He $A$-$B$
mixtures~\cite{salomaa_nature_1987,volovik,finne_rpp_2006}
and in Early-Universe cosmology. It has been
proposed that a series
of symmetry breakings lead to formation of domain walls and cosmic
strings, which
terminate on the boundaries between regions of different vacuum
states~\cite{kibble_jpa_1976,vilenkin-shellard}.
Highly complex interface physics
also arises from collisions between branes~\cite{sarangi_plb_2002} in
string-theory brane-inflation~\cite{dvali_plb_1999} scenarios.
Superfluids have been discussed as candidates for experimentally accessible
systems where analogues of cosmic topological defects may be
studied~\cite{zurek_nature_1985,volovik}.  For example, colliding liquid $^3$He
$A$-$B$ interfaces
have been proposed as analogues of string-theory
branes~\cite{bradley_nphys_2008}. 

Here we show how atomic-physics laboratory techniques can be employed
for engineering topologically nontrivial, coherent interface
boundaries between spatially separated different ground-state
manifolds which simultaneously exhibit different broken symmetries. We
demonstrate nontrivial penetration of singular defects across
a constructed stable interface between ferromagnetic (FM)
and polar phases of a spin-1 BEC.
We identify the basic defect solutions crossing the interface and
minimize their energies in order to characterize the defect core
structures. We find examples of intriguing core deformations where
a singular vortex terminates as an arch defect on the interface with
the topological charge of a monopole, and where a coreless,
nonsingular vortex connects to a pair of singular half-quantum
vortices.

Our example demonstrates how the ultracold
atom interface physics provides a novel medium for studies of stability
properties of field-theoretical
solitons~\cite{bogomolny_sjnp_1976,jackiw_prd_1976,manton_sutcliffe,faddeev_nature_1997}.
The spin-1 BEC also already provides a possible system for dynamical
investigation of phase transitions and
defect production, e.g., of brane annihilation models. Moreover, the
proposed method for interface engineering can exhibit especially rich
phenomenology in spin-2 and spin-3 BECs and in strongly correlated
optical lattice systems. 

Atomic-physics technology provides tools for accurate detection methods for
ultracold-atom systems on length and time scales that are difficult
to achieve in laboratory systems of more traditional quantum fluids and
solids. Advanced measurement techniques combined with the high degree
of control over experimental parameters make them suitable for quantum
simulators of physical phenomena that are too complex even for numerical
studies. This has attracted considerable interest, in particular in
optical lattice systems, which can emulate strongly correlated
condensed-matter models.
The experimental development for using ultracold atoms as a laboratory
testing ground for complex physical phenomena has been accelerated,
e.g., by the observations of the Mott-insulator states of
atoms~\cite{greiner_nature_2002,jordens_nature_2008,schneider_science_2008},
the study of nonequilibrium defect formation in phase transitions in the
Kibble-Zurek mechanism for both scalar~\cite{weiler_nature_2008} and
spin-1 BECs~\cite{sadler_nature_2006},
and in the preparation of artificial gauge-field potentials for multi-level
atoms~\cite{lin_nature_2009}.

Spinor BECs are condensates in which the spin degree of freedom is not
frozen by magnetic trapping~\cite{stenger_nature_1998}. They provide
ideal models and emulators of complex broken symmetries due to their rich
phenomenology of different
phases~\cite{ho_prl_1998,ohmi_jpsj_1998,zhou_ijmpb_2003,koashi_prl_2000,ciobanu_pra_2000,barnett_prl_2006,santos_spin-3_2006}
that support exotic defects and
textures~\cite{stoof_monopoles_2001,ruostekoski_monopole_2003,semenoff_prl_2007,kobayashi_prl_2009}.
Spinor BECs have attracted recent experimental
attention, e.g., in the studies of formation of spin
textures~\cite{vengalattore_prl_2008,kronjager_prl_2010} and in controlled
preparation~\cite{leanhardt_prl_2003,leslie_prl_2009} and nonequilibrium
formation~\cite{sadler_nature_2006} of vortices.

Here we will concentrate on a spin-1 BEC whose
macroscopic wave function $\Psi({\bf r})$ may be
written in terms of the local
density $n({\bf r})$ and a normalized spinor
$\zeta({\bf r})$ as,
\begin{equation}
  \label{eq:spinor}
  \Psi({\bf r}) = \sqrt{n({\bf r})}\zeta({\bf r})
  = \sqrt{n({\bf r})}\threevec{\zeta_+({\bf r})}
                              {\zeta_0({\bf r})}
			      {\zeta_{-}({\bf r})},
  \quad
  \zeta^\dagger\zeta=1.
\end{equation}
In the Gross-Pitaevskii mean-field description, the Hamiltonian density of the
spin-1 BEC reads
\begin{equation}
  \label{eq:hamiltonian-density}
  \begin{split}
    {\cal H} =  &\frac{\hbar^2}{2m}\abs{\nabla\Psi}^2 + V(\bvec{r})n
    + \frac{c_0}{2}n^2
    + \frac{c_2}{2}n^2\abs{\bvec{\eva{\hat{F}}}}^2 \\
    &+ g_1n\eva{\bvec{B}\cdot\bvec{\hat{F}}}
    + g_2n\eva{\left(\bvec{B}\cdot\bvec{\hat{F}}\right)^2}\,.
  \end{split}
\end{equation}
$\bvec{\inleva{\hat{F}}}=\zeta_\alpha^\dagger\bvec{\hat{F}}_{\alpha\beta}\zeta_\beta$
is the expectation value of the spin operator $\bvec{\hat{F}}$, defined as a vector of
spin-1 Pauli matrices.  A weak external magnetic field leads to linear
and quadratic Zeeman shifts described by the last two terms.
The two interaction strengths are $c_0 = 4\pi\hbar^2(2a_2+a_0)/3m$ and
$c_2 = 4\pi\hbar^2(a_2-a_0)/3m$, respectively, where $m$ is the atomic
mass, and $a_0$ and $a_2$ are the $s$-wave scattering lengths
corresponding to the two different
values of the relative angular momentum of the colliding atom pair.

The sign of $c_2$ determines which phase is energetically favored by
the interaction alone. For $c_2<0$, as with $^{87}$Rb, minimization of
the interaction
energy favors the FM phase with the maximum spin magnitude
$|\langle\bvec{\hat{F}}\rangle|=1$  in which case the broken symmetry
of the ground-state manifold
is defined by the rotations of the spin vector.
The FM phase supports two topologically distinct
classes of line defects~\cite{ho_prl_1998,ohmi_jpsj_1998}.  The
nontrivial vortices 
in each class are singly quantized singular line defects and
nonsingular coreless vortices, respectively~\footnote{See
the Appendix for basic defect configurations in spin-1 BECs and the
construction of prototype spinor wave functions for vortex connections
across the interface.}, both of which have been
observed~\cite{sadler_nature_2006,leanhardt_prl_2003,vengalattore_prl_2008,leslie_prl_2009}.

If instead $c_2>0$, as with $^{23}$Na, the interaction energy favors
the polar phase, minimizing
the spin magnitude $|\langle\bvec{\hat{F}}\rangle|=0$.
The broken symmetry of the ground-state manifold is described by the
unoriented nematic axis $\nematic$ ($\nematic=-\nematic$) and the
condensate phase $\phi$.
The polar phase therefore exhibits nematic order, analogously to liquid
crystals and the superfluid liquid $^3$He-$A$ phase, and so supports
both integer  and half-quantum vortices.

Scattering lengths in ultracold-atom systems are routinely manipulated
using magnetic Feshbach resonances.  However, this is not possible in
a spinor BEC, since the required strong magnetic field would freeze out the
condensate spin degree of freedom.  It is possible to manipulate
scattering lengths also
through
optical~\cite{fatemi_prl_2000} or
microwave-induced Feshbach
resonances~\cite{papoular_pra_2010} in which case the fields can be
kept sufficiently
weak in order not to destroy the spinor nature of the BEC.  The
Feshbach resonance changes the scattering length by coupling the
entrance channel to a virtually populated bound
state~\cite{papoular_pra_2010}. In particular, it is possible to
tune the ratio $a_0/a_2$ between the two scattering lengths.

We suggest constructing
an interface between topologically distinct manifolds by
spatially nonuniform
engineering of the scattering lengths. In a spin-1 BEC this may be
experimentally
realized to prepare an interface between coexisting FM and
polar phases. Using an optical
\mbox{Feshbach} resonance, the spatial pattern corresponding to a sharp
interface can be imposed using a
holographic mask. The spin-dependent
interaction strength $c_2$ is proportional to the difference between
$a_2$ and $a_0$. Thus for small $|c_2|$, as is the case with both
$^{87}$Rb and $^{23}$Na, only a small relative shift
of $a_0$ versus $a_2$ is necessary to prepare the interface, and
therefore the inelastic
losses associated with optical Feshbach
resonances~\cite{fatemi_prl_2000} can be kept small.

A microwave field cannot similarly be focused. However,
using an optically induced level
shift to tune the microwave transition off-resonant where no
adjustment of the scattering length is desired, interactions may be
manipulated in spatially well-defined regions to prepare a sharp
interface boundary
without the losses associated with the optical Feshbach resonance.

In order to demonstrate the nontrivial nature of defect penetration
across an interface
between topologically distinct manifolds, we consider a harmonically
trapped spin-1 BEC, where $c_2$ abruptly changes sign at the center of
the trap at $z=0$. We have $c_2>0$ for $z>0$, corresponding to the polar
phase, and for $z<0$, the BEC is in the FM phase with
$c_2<0$.

\begin{figure*}[tbp]
  \centering
  \includegraphics[width=\textwidth]{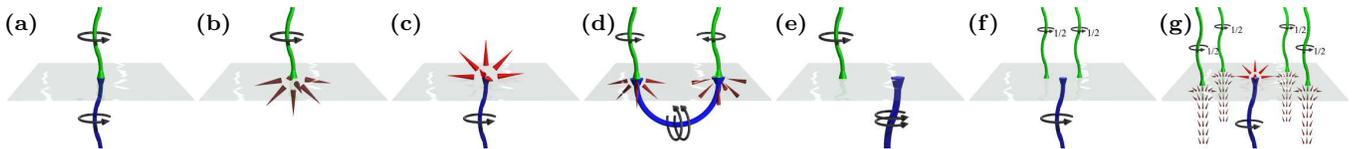}
  \caption{(color online).
    Schematic illustrations of possible vortex connections. The polar phase
    is above the interface and the FM phase is below.
    (a) A singly quantized vortex in both phases.
    (b) A
    singly quantized vortex in the polar phase can connect to a Dirac
    monopole. The Dirac monopole can be continuously transformed into a
    coreless vortex.
    (c) A singly quantized spin vortex terminates as a polar monopole.
    (d) A dipole can be constructed by joining the Dirac strings of a Dirac
    monopole and an antimonopole~\cite{savage_dirac_2003}.  Placed on
    the interface, the dipole
    connects to two singly quantized vortices on the polar side.
    (e) A singly quantized vortex in the polar phase connecting to a
    doubly quantized vortex on the FM side may be cut in half at the
    interface and the resulting vortices
    in the two regions may be moved apart if
    an additional dark-soliton plane is introduced in $\zeta_0$.
    More complicated vortex states may form as splitting of a singly quantized
    polar vortex into two half-quantum vortices (f), or by
    nucleation of half-quantum vortices that connect to coreless
    vortices that may exist together with monopoles (g).}
  \label{fig:schematic}
\end{figure*}
We first construct spinor solutions that approximate physical
wave functions simultaneously in the two different manifolds and
quickly relax to
vortices connecting across the interface or terminating at the interface.
Some illustrative examples of topologically allowed states are shown
in Fig.~\ref{fig:schematic}.
The simplest connection can be identified by considering a
singly quantized vortex in both phases. Although
such a vortex represents a very different topology in the two phases,
it can be formed in both cases, e.g., by a $2\pi$ winding of the
condensate phase $\phi$ around the $z$ axis. The two vortex solutions
can be joined
by changing the sign of either $\zeta_+$ or $\zeta_-$. By appropriate
choice of parameters,  doing so
causes the spinor wave function to adjust between the two manifolds by
forcing $\absF$ to switch from 0 to 1
[Fig.~\ref{fig:schematic}(a)], or else leads to a state which
  immediately relaxes to the desired configuration.
Physically, such a sign change in one
of the two spinor components can be obtained by introducing a dark
soliton plane (phase kink) in that component at $z=0$, in which case
the $\pi$-phase shift across the soliton is associated with a
vanishing density at the soliton core. As the density of the other
components at the soliton core does not vanish, the BEC wave function
continuously connects the two manifolds. The interface acquires a
width determined by the spin healing length
$\xi_F=(8\pi |c_2| n)^{-1/2}$, the length scale over
which $|\langle\bvec{\hat{F}}\rangle|$ heals when locally
perturbed.

A singular vortex with unit winding in the polar phase can also be
written as a $2\pi$
spin rotation about the $z$ axis together with a $2\pi$ rotation of
the condensate phase $\phi$.
If we continue this solution to the FM side, by
changing the sign of $\zeta_-$, we identify the resulting structure as
an approximation of a coreless
vortex~\cite{ho_prl_1998}, in which the spin profile quickly acquires a
fountain-like texture. Hence we have constructed a solution where a
polar, singly quantized vortex connects to a FM coreless
vortex [Fig.~\ref{fig:schematic}].  We parametrize the spinor as
\begin{equation}
  \label{eq:012}
  \zeta^{\rm 1 \leftrightarrow cl} = \frac{1}{\sqrt{2}}\threevec{-\sin\beta}
	                             {\sqrt{2}e^{i\varphi}\cos\beta}
				     {\pm e^{2i\varphi}\sin\beta},
\end{equation}
where $\varphi$ is the azimuthal coordinate, and $\beta=3\pi/4$ gives
an exact switch from polar to FM, with the
negative sign in $\zeta_-$ used in the FM region.

If the coreless vortex in Eq.~(\ref{eq:012}) is continuously deformed
into a doubly
quantized, singular vortex along the positive $z$ axis terminating at
the origin, the resulting
spin texture on the FM side forms a radial hedgehog,
$\inleva{\bvec{\hat{F}}}=\bvec{\hat{r}}$. This structure
can be identified as the analogue of the Dirac monopole in quantum
field theory, with the singular vortex line representing the Dirac
string~\cite{savage_dirac_2003} [Fig.~\ref{fig:schematic}(b)]. The
deformation is possible due to topological equivalence
between the doubly quantized vortex and the vortex-free state.

A singular spin vortex in the FM phase can be made to terminate on a
polar monopole [Fig.~\ref{fig:schematic}(c)] as follows: The polar
monopole is formed by two overlapping vortex lines of opposite circulation
in $\zeta_\pm$ perpendicular to a soliton plane in $\zeta_0$. The nematic
axis then exhibits the radial
hedgehog~\cite{stoof_monopoles_2001,ruostekoski_monopole_2003}
$\nematic=\bvec{\hat{r}}$, which is the analogue of the t'Hooft-Polyakov
monopole. Inserting a phase kink in $\zeta_+$ at the interface results in a
structure on the FM side where $\inleva{\bvec{\hat{F}}}$ points
radially away from the $z$ axis, which we identify as the spin vortex,
\begin{equation*}
  \zeta^{\rm sv\leftrightarrow pm} =
  \frac{1}{\sqrt{2}}\threevec{\mp e^{-i\varphi}\sin\theta}
                             {\sqrt{2}\cos\theta}
	                     {e^{i\varphi}\sin\theta},
\end{equation*}
where $\theta$ and $\varphi$ denote the spherical angles, and the positive
sign is used in the FM part.

\begin{figure}[tbp]
  \centering
  \includegraphics[scale=1]{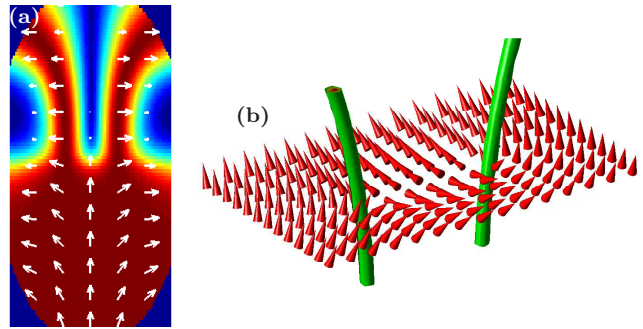}
\caption{(color online). Core
  structure after minimizing the energy of Eq.~(\ref{eq:012}),
  corresponding to Fig.~\ref{fig:schematic}(b).
  (a) The magnitude of the spin [$\absF=1$ is
  dark red (dark gray) with long arrows]. The polar vortex has split
  into two half-quantum vortices 
  with FM cores with nonvanishing densities. White arrows show the
  spin vector and
  indicate the coreless vortex in the FM
  part.  Here $c_0=2.0\times10^4\hbar\omega l^{3}$,
  $\abs{c_2}=2.5\times10^2\hbar\omega l^{3}$, $\Omega=0.12\omega$ and $B=0$, where $l=(\hbar/m\omega)^{1/2}$ is the oscillator length (for $^{87}$Rb with $\omega=2\pi\times10\,$Hz this would correspond to $10^6$ atoms).
  (b) The
  nematic axis $\nematic$ (unoriented but shown as cones to emphasize winding)
  displays the characteristic $\pi$ winding as each half-quantum vortex is
  encircled. The two cores are joined by a disclination plane
  indicating the turn of $\nematic$ by
  $\pi$. ($\abs{c_2}=1.0\times10^4\hbar\omega l^{3}$, $\Omega=0.19\omega$.)
}
\label{fig:012}
\end{figure}
\begin{figure*}[tbp]
  \centering
  \includegraphics[width=\textwidth]{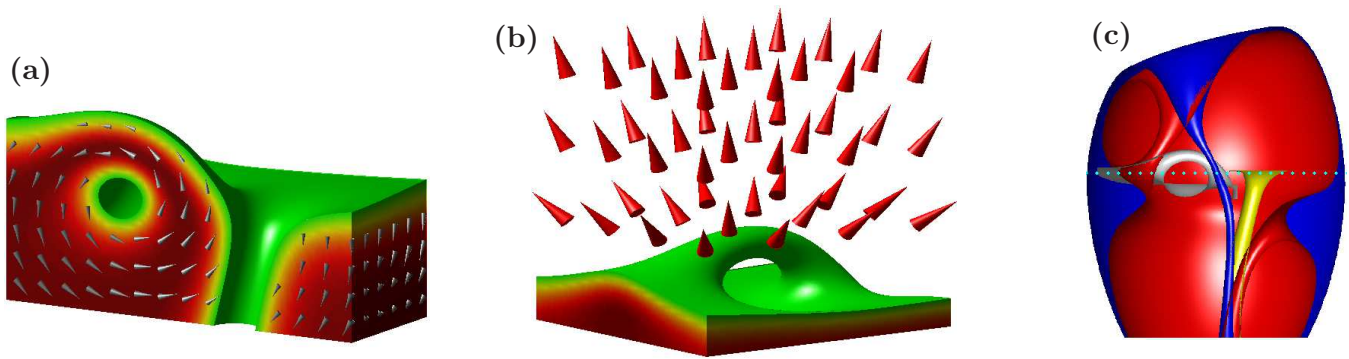}
\caption{(color online). Alice arch. When a singly quantized spin vortex
  terminates on a polar monopole, the point defect deforms into an
  arch-shaped line defect. (a) An isosurface of the spin
  magnitude is shown in green (light gray).  The spin magnitude rises
  to $1$ [dark red (black)] on the FM side of the interface ($z<0$) and
  inside the Alice
  arch on the polar side.  The singly quantized vortex with polar core
  remains in the
  FM phase. Gray cones indicate the spin vector.
  (b) The nematic axis (cones) retains the hedgehog structure centered
  on the Alice arch (the deformed point defect), indicating that the
  topological
  charge is preserved.
  (c) Constant-density surfaces for $n\abs{\zeta_+}^2$ [red (medium
  gray)] and $n\abs{\zeta_-}^2$ [blue (dark gray)].
  The Alice arch is formed by deformation
  of the vortex cores in the $\zeta_\pm$
  spinor components. The vortex line in $\zeta_+$
  splits at the interface (indicated by the
  dotted line). The upper part forms
  the Alice arch together with the vortex line in $\zeta_-$ above the
  interface. The arch (above the interface) and the spin vortex (below the interface) 
  are indicated by silver and gold (light gray) spin
  isosurfaces at $\absF=0.9$ and $\absF=0.5$,
  respectively. ($c_0=2.0\times10^4\hbar\omega l^{3}$,
  $\abs{c_2}=5.0\times10^2\hbar\omega l^{3}$, $\Omega=0$ and $B=0$.)}
\label{fig:alice}
\end{figure*}
So far we have analyzed the topological existence of defects
perforating the FM-polar interface. In
order to determine their energetic stability and the core structure we
numerically minimize the energy of the constructed defect in a rotating frame
$F = E-\Omega \langle\hat{L}_z\rangle$ by evolving the corresponding
coupled spin-1 Gross-Pitaevskii equations in imaginary time. Here
$\Omega$ denotes the frequency of
rotation that is assumed to be around the $z$ axis,
$\langle\hat{L}_z\rangle$ is the
$z$-component of the angular momentum and
$E=\int d^3r {\cal H}({\bf r})$.
We assume a slightly cigar-shaped trap
$\omega_x=\omega_y=2\omega_z\equiv \omega$~\footnote{The
  results are qualitatively similar in the presence of a weak Zeeman
  splitting ($|g_1 {\bf B}| \le 10^{-1}\hbar\omega$) to the case of ${\bf B}=0$.
  We also checked that fixing the magnetization
  $M=N_{+}-N_{-}$ (where $N_\pm$ are the total populations of
  $\zeta_\pm$) during the relaxation does not alter the basic
  results.}.

Minimizing the the energy of $\zeta^{\rm 1\leftrightarrow cl}$
results in the core
deformation shown in Fig.~\ref{fig:012}. The state exhibits the
coreless vortex on the FM side of the interface
[Fig.~\ref{fig:012}(a)]. The frequency of rotation determines the
direction of the spin vector at the edge of the cloud. The core of the
singly quantized vortex in the polar part is
deformed into two half-quantum vortices.

The size of a singular defect core with vanishing density is determined
by $\xi_n=(8\pi c_0 n)^{-1/2}$, the density healing length. The
singular polar vortex lowers its energy by spontaneously breaking
axial symmetry, splitting into half-quantum vortices with FM
cores of size $\xi_F$. The deformation is
energetically favorable when the energy cost of the FM
cores is smaller than the energy gained by removing the density
depletion. We find the same splitting of the polar vortex for the
defect with a singular vortex in both phases.

A very intriguing core deformation results from minimizing the energy of
a singular spin vortex terminating on a
polar monopole [Fig.~\ref{fig:schematic}(c)].
The point defect requires the density to go to zero.
For sufficiently large $\xi_F$, the energy cost of the density
depletion can be avoided by deforming
the point defect into a semicircular line defect with FM
core whose ends attach to the interface.  Figure~\ref{fig:alice} shows
the resulting arch-like defect.

The arch is formed as a local deformation of the point defect,
and retains its topological charge. Specifically, the radial hedgehog
in the nematic axis $\nematic$ is preserved away from the arch.
Single-valuedness of $\Psi$ then requires that $\nematic$ turn by $\pi$ on
any closed loop through the arch, accompanied by a $\pi$ change in the
phase $\phi$.  We thus identify the line defect as an arch-shaped
half-quantum vortex, which we will call an Alice arch, as it is
the interface analogue of the complete Alice
ring~\cite{ruostekoski_monopole_2003}. Such ring-shaped defects are analogous
to Alice rings in high energy physics~\cite{schwarz_npb_1982}
with a topological charge
similar to the magnetic ``Cheshire" charge~\cite{alford_npb_1991}
We find that the deformation of a point defect to an arch is energetically
favorable for $c_2 \lesssim 0.5c_0$. It is unstable towards drifting
out of the cloud due to the density gradient, but could be stabilized
by a weak pinning potential.

Different defect structures penetrating the interface can be
prepared experimentally by recognizing that the defects are composed
of simple vortex lines, phase kink planes or, in more complex cases,
by vortex rings in the three spinor components, each of which may be
phase-imprinted using existing
technologies~\cite{matthews_prl_1999,ruostekoski_prl_2001,ruostekoski_pra_2005}.
Vortices may also nucleate
due to rotation.  We find that nucleation energetically favors defects
consisting of a half-quantum vortex connecting to a
coreless vortex, leading to states such as that illustrated in
Fig.~\ref{fig:schematic}(g).

Here we have demonstrated topological interface engineering by
studying examples of defect perforation across a FM-polar interface in
a spin-1 BEC. 
Vortex bifurcation purely due to an energetic (not topological) effect
in the phase separation of a two-species BEC was studied in
Refs.~\cite{takeuchi_jpsj_2006, kasamatsu_jhep_2010}.
Our method can be extended to more complex broken symmetries in
\mbox{spin-2}~\cite{koashi_prl_2000,ciobanu_pra_2000,semenoff_prl_2007} and
\mbox{spin-3}~\cite{barnett_prl_2006,santos_spin-3_2006} BECs that also
support, e.g., non-Abelian
defects~\cite{kobayashi_prl_2009,huhtamaki_pra_2009}. Other
particularly promising systems for topological interface physics are
strongly correlated atoms in optical
lattices~\cite{greiner_nature_2002,jordens_nature_2008,schneider_science_2008}
exhibiting also quantum phase transitions and potential analogues of exotic
superconductivity~\cite{bert_nphys_2011} in crystal lattices.

Nonequilibrium defect production may be investigated in phase
transitions in the presence of different broken
symmetries~\cite{kibble_jpa_1976,zurek_nature_1985}. Defect formation
from colliding interfaces can be employed as a model to simulate
cosmological brane
annihilation~\cite{dvali_plb_1999,sarangi_plb_2002}.
For instance, in a FM condensate, a slab of polar phase could be
created, each interface being a 2D analogue of a $D$-brane. Removing the
interaction shift causes the slab to collapse, bringing the interfaces
closer until they meet and annihilate, representing an annihilation of
a brane-antibrane pair. In braneworld scenarios of cosmic inflation
the annihilation may lead to defect production~\cite{sarangi_plb_2002}
that could be directly observed in atomic BECs. A similar experiment
has been performed with superfluid liquid $^3$He $A$-$B$ interfaces in which
case, however, the detection of defects is
difficult~\cite{bradley_nphys_2008}.

We acknowledge discussions with D.~J.\ Papoular and M.~D.\ Lee and
financial support from Leverhulme Trust.

\appendix

\setcounter{equation}{0}
\setcounter{figure}{0}
\renewcommand{\theequation}{A\arabic{equation}}
\renewcommand{\thefigure}{A\arabic{figure}}

\section{Appendix}

In this Appendix we present the basic defect configurations of spin-1
Bose-Einstein condensate (BEC) corresponding to the different broken
symmetries of the ferromagnetic (FM) and polar phases. 
Using these solutions we construct prototype spinor wave functions for
the vortex connections across the interface between the polar-FM
manifolds (such as those shown 
in Fig. 1 of the main text). The constructed prototype defect
configurations are used as initial states for the numerical studies of
the defect stability in the main text. 

\subsection*{Broken symmetries and vortices in the spin-1
  Bose-Einstein Condensate}

The wave function of a spin-1 BEC is written in terms of the density
$n(\bvec{r}) = |\Psi(\bvec{r})|^2$ and a normalized three-component spinor in
the basis of spin projection onto the $z$ axis
\begin{equation}
  \label{eq:spinor_suppl}
  \Psi({\bf r}) = \sqrt{n({\bf r})}\zeta({\bf r})
  = \sqrt{n({\bf r})}\threevec{\zeta_+({\bf r})}
                              {\zeta_0({\bf r})}
                              {\zeta_{-}({\bf r})}.
\end{equation}
The Hamiltonian density of a BEC of spin-1 atoms
is~\cite{pethick-smith}
\begin{equation}
  \label{eq:hamiltonian-density-app}
  \begin{split}
    {\cal H} &=  \frac{\hbar^2}{2m}\abs{\nabla\Psi}^2 + V(\bvec{r})n
    + \frac{c_0}{2}n^2
    + \frac{c_2}{2}n^2\abs{\bvec{\eva{\hat{F}}}}^2\\
    &+ g_1n\eva{\bvec{B}\cdot\bvec{\hat{F}}}
    + g_2n\eva{\left(\bvec{B}\cdot\bvec{\hat{F}}\right)^2},
  \end{split}
\end{equation}
where $V(\bvec{r})$ denotes the trapping potential, $m$ is the atomic
mass, $\bvec{\hat{F}}$
is the spin operator, given by
a vector of spin-1 Pauli matrices, and
$\bvec{\inleva{\hat{F}}}=\zeta_\alpha^\dagger\bvec{\hat{F}}_{\alpha\beta}\zeta_\beta$.
If a weak, external magnetic
field is present, this will lead to linear and quadratic Zeeman shifts
described by the last two terms.

There are two
interaction terms with strengths
$c_0=4\pi\hbar^2(2a_2+a_0)/3m$ and $c_2=4\pi\hbar^2(a_2-a_0)/3m$,
where $a_{\!f}$ is the s-wave scattering length in the scattering channel
with relative angular momentum $f$.  In $^{23}$Na,
$c_0/c_2\simeq 31$~\cite{crubellier_epjd_1999} and in $^{87}$Rb 
$c_0/c_2\simeq -216$~\cite{van-kempen_prl_2002}.  The two
corresponding healing lengths 
in the spin-1 BEC, $\xi_n=(8\pi c_0 n)^{-1/2}$ and
$\xi_F=(8\pi |c_2| n)^{-1/2}$,  define the length scales over
which $n$ and $|\langle\bvec{\hat{F}}\rangle|$ heal when locally
perturbed.

The spin-dependent interaction in spin-1 BEC gives rise to two phases, the FM and polar~\cite{pethick-smith,ho_prl_1998,ohmi_jpsj_1998,leonhardt_jetplett_2000,zhou_ijmpb_2003,ruostekoski_monopole_2003,savage_dirac_2003}.
Both phases exhibit distinct sets of possible defects.
When a defect core is formed by a depletion of
the density, its size is determined by the density healing length
$\xi_n$.  A defect core in the polar (FM) phase may instead be filled
with atoms in the FM (polar) phase. The size of the core is then
determined by the spin healing length $\xi_F$~\cite{ruostekoski_monopole_2003}.
In the following we present the basic defect structures in the FM and polar
phases of a spin-1 BEC that are used in constructing the defect solutions
crossing the coherent topological interface between the FM and polar phases.

{\em FM phase:}
If $c_2<0$, the spin-dependent interaction
will favor a state that maximizes the magnitude of the spin, i.e.,
$\absF=1$.  The general spinor may then be written
as spin-rotations
of $\zeta = (1,0,0)^T$, together with a condensate phase $\phi$
\begin{equation}
  \label{eq:ferro}
  \zeta^{\rm f} =
  e^{i\phi}U(\alpha,\beta,\gamma)\threevec{1}{0}{0}
  = \frac{e^{-i\gamma^\prime}}{\sqrt{2}}
    \threevec{\sqrt{2}e^{-i\alpha}\cos^2\frac{\beta}{2}}
             {\sin\beta}
             {\sqrt{2}e^{i\alpha}\sin^2\frac{\beta}{2}},
\end{equation}
where $(\alpha,\beta,\gamma)$ are spin-space Euler angles, and
\mbox{$\gamma^\prime=\gamma-\phi$} absorbs the condensate phase. The
broken symmetry of the ground-state manifold therefore corresponds to the
group of three-dimensional rotations \SO(3). The spin vector is given
by the Euler angles as
\mbox{$\inleva{\bvec{\hat{F}}}=(\cos\alpha\sin\beta,\sin\alpha\sin\beta,\cos\beta)$.}

\begin{figure}[tb]
  \centering
  \includegraphics[scale=1]{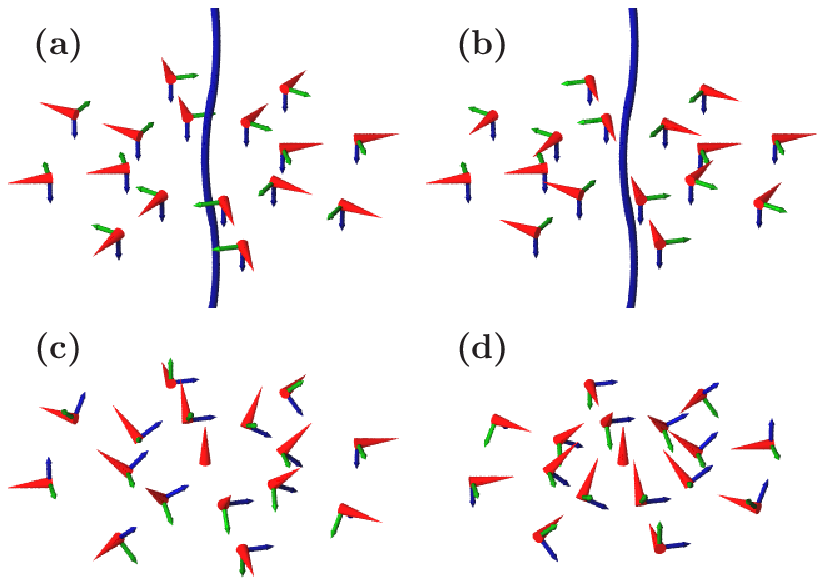}
\caption{Nontrivial vortices in the FM phase.
  (a) The singular spin vortex is formed as a radial disgyration of
  the spin vector (red cones) around the singular core.
  (b) Singular spin vortex with cross-disgyration of the spin vector.
  (c) The coreless vortex is
  formed as a combined disgyration of the spin vector and a winding of
  the condensate phase, corresponding to spin rotations about the
  local spin vector (indicated by the orthogonal green and blue
  vectors). The core is
  nonsingular and filled by the vortex-free spinor component.
  The direction of the spin vector at the edge depends on
  boundary conditions.
  (d) A different nonsingular vortex configuration.
}
\label{fig:fmVortices}
\end{figure}

Topological stability of line defects can be characterized by studying
closed contours around the defect line and the mapping of these contours to the
order parameter space~\cite{mermin_rmp_1979}. If the image of a closed
loop encircling a line defect
in the order parameter space can be contracted to a point, the defect
is not topologically
stable.  $\SO(3)$ can be represented geometrically as $S^3$ (the unit
sphere in four dimensions) with diametrically opposite points
identified. The only closed contours that cannot be contracted
to a point are then the ones connecting these diametrically opposite
points, and we have only two distinct classes of vortices: singly
quantized, singular vortices that correspond to noncontractible loops
and nonsingular vortices representing contractible loops. All other
vortices can be transformed to either one of these by local
deformations of the order parameter. Examples of this first class are
a singular line defect constructed as a winding of the condensate
phase,
\begin{equation}
  \label{eq:fmsingular}
  \zeta^{\rm s} =
  \frac{e^{i\varphi}}{\sqrt{2}}\threevec{\sqrt{2}\cos^2\frac{\beta}{2}}
                                       {\sin\beta}
                                       {\sqrt{2}\sin^2\frac{\beta}{2}},
\end{equation}
where $\varphi$ is the azimuthal angle, and the singular spin
vortex constructed as a $2\pi$ rotation of the spin
vector~\cite{ho_prl_1998,ohmi_jpsj_1998} [Fig.~\ref{fig:fmVortices}(a) and (b)]
\begin{equation}
  \label{eq:spinvortex}
  \zeta^{\rm sv} =
  \frac{1}{\sqrt{2}}\threevec{\sqrt{2}e^{-i\varphi}\cos^2\frac{\beta}{2}}
                             {\sin\beta}
                             {\sqrt{2}e^{i\varphi}\sin^2\frac{\beta}{2}}.
\end{equation}
A nonsingular, coreless
vortex is constructed as a combined rotation of the spin vector and
the condensate phase [Fig.~\ref{fig:fmVortices}(c) and (d)]
\begin{equation}
  \label{eq:cl}
  \zeta^{\rm cl} =
  \frac{1}{\sqrt{2}}\threevec{\sqrt{2}\cos^2\frac{\beta}{2}}
                             {e^{i\varphi}\sin\beta}
                             {\sqrt{2}e^{2i\varphi}\sin^2\frac{\beta}{2}}.
\end{equation}

\begin{figure}[tb]
  \centering
  \includegraphics[scale=1]{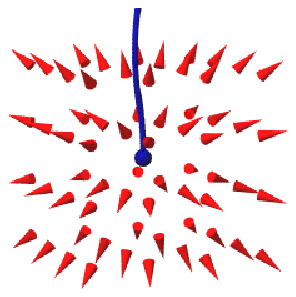}
\caption{Dirac monopole; the analogue of the Dirac monopole is
  formed as the termination of a doubly quantized, singular vortex
  line (blue) resulting in a hedgehog configuration of the spin vector.
}
\label{fig:dirac}
\end{figure}

$\SO(3)$ does not strictly speaking support point defects.
However, it is possible to form a spinor with a
monopole structure of the spin vector (a radial hedgehog) as the
termination of a doubly quantized vortex~\cite{savage_dirac_2003}
[Fig.~\ref{fig:dirac}]
This is the analogue of the Dirac monopole in quantum field theory.
\begin{equation}
  \label{eq:dirac}
  \zeta^{\rm D} = \frac{1}{\sqrt{2}}
                  \threevec{\sqrt{2}e^{-2i\varphi}\cos^2\frac{\theta}{2}}
                   {e^{-i\varphi}\sin\theta}
                   {\sqrt{2}\sin^2\frac{\theta}{2}}.
\end{equation}

{\em Polar phase:} $c_2>0$ favors a state with $\absF=0$. The general
spinor is given by spin rotations of $\zeta=(0,1,0)^T$ together with
a condensate phase
\begin{equation}
  \label{eq:polar}
  \zeta^{\rm p} =
  e^{i\phi}U(\alpha,\beta,\gamma)\threevec{0}{1}{0}
  = \frac{e^{i\phi}}{\sqrt{2}}\threevec{-e^{-i\alpha}\sin\beta}
                     {\sqrt{2}\cos\beta}
                     {e^{i\alpha}\sin\beta}.
\end{equation}
The unit vector
$\nematic = (\cos\alpha\sin\beta,\sin\alpha\sin\beta,\cos\beta)$
defines the local direction of macroscopic condensate spin
quantization.  We may then write the spinor as
\begin{equation}
  \zeta^{\rm p} = \frac{e^{i\phi}}{\sqrt{2}}
                 \threevec{-d_x+id_y}{\sqrt{2}d_z}{d_x+id_y}.
\end{equation}
Note, however, that
$\zeta^{\rm p}(\phi,\nematic) = \zeta^{\rm p}(\phi+\pi,-\nematic)$.
These two states must be identified in order to avoid double counting.
The order parameter space, which might appear to be $\U(1)\times S^2$,
from the condensate phase and the rotations of $\nematic$, actually becomes
$(\U(1) \times S^2)/\mathbb{Z}_2$, where $\mathbb{Z}_2$ denotes a
discrete two-element group.
The vector $\nematic$ is taken to be {\em unoriented\/} and defines
the {\em nematic axis,} which together with $\phi$ fully specifies the
broken symmetry.

\begin{figure}[tb]
  \centering
  \includegraphics[scale=1]{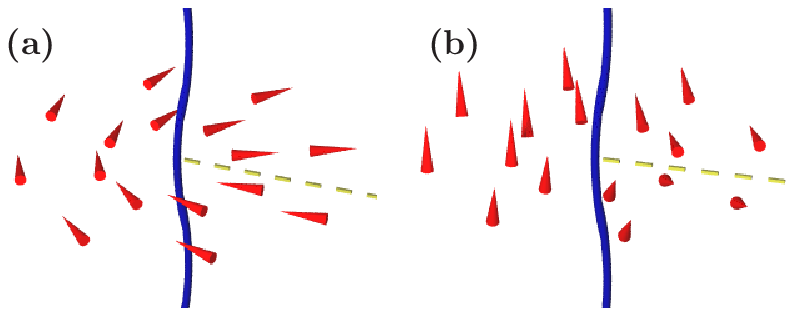}
\caption{Half-quantum vortex.
  (a) In the half-quantum vortex described by Eq.~(\ref{eq:hq})
  the nematic axis (red cones) winds by $\pi$ as the core is
  encircled. This is accompanied by a $\pi$ winding of the condensate
  phase. The disclination plane is indicated by the dashed line.
  (b) Another possible $\pi$ winding of the nematic axis in a
  half-quantum vortex.
}
\label{fig:hq}
\end{figure}

A singly quantized vortex can again be constructed as a $2\pi$ winding
of the condensate phase
\begin{equation}
  \label{eq:psingular}
  \zeta^{\rm 1} =
  \frac{e^{i\varphi}}{\sqrt{2}}\threevec{-e^{-i\alpha}\sin\beta}
                     {\sqrt{2}\cos\beta}
                     {e^{i\alpha}\sin\beta}.
\end{equation}
The phase winding may be combined with a $2\pi$ winding of the nematic
axis by choosing $\alpha=\varphi$
\begin{equation}
  \label{eq:p012}
  \zeta^{\rm 1^\prime} =
  \frac{1}{\sqrt{2}}\threevec{-\sin\beta}
                             {\sqrt{2}e^{i\varphi}\cos\beta}
                             {e^{2i\varphi}\sin\beta}.
\end{equation}
Because of the property that a $\pi$ change in $\phi$ can be exactly
compensated by a $\pi$ rotation of $\nematic$, it is possible to form
vortices carrying half a quantum of circulation by letting both $\phi$
and $\nematic$ wind by $\pi$ as the vortex is
encircled~\cite{leonhardt_jetplett_2000}, leading to a
disclination plane where $\nematic\rightarrow-\nematic$.  If $\nematic$
is in the $(x,y)$-plane, a half-quantum vortex can be
written
\begin{equation}
  \label{eq:hq}
  \zeta^{\rm hq} = \frac{e^{i\varphi/2}}{\sqrt{2}}
                   \threevec{-e^{-i\varphi/2}}
                            {0}
                            {e^{i\varphi/2}}
		 = \frac{1}{\sqrt{2}}
                   \threevec{-1}
                            {0}
                            {e^{i\varphi}}.
\end{equation}
In general, the axis about which $\nematic$ winds need not coincide
with the vortex core.  The $\pi$ winding of the nematic axis still
allows us to identify the vortex [Fig.~\ref{fig:hq}].

The polar phase supports also point defects~\cite{stoof_monopoles_2001,ruostekoski_monopole_2003}.
We can form a radial hedgehog of the nematic axis by choosing
$\nematic = \bvec{\hat{r}} =
(\sin\theta\cos\varphi,\sin\theta\sin\varphi,\cos\theta)$,
which represents a spherically symmetric monopole,
\begin{equation}
  \label{eq:polar_monopole}
  \zeta^{\rm pm} = \frac{1}{\sqrt{2}}\threevec{-e^{-i\varphi}\sin\theta}
                            {\sqrt{2}\cos\theta}
	                    {e^{i\varphi}\sin\theta}.
\end{equation}
Here the origin corresponds to a singular point defect. The two spinor
wave function components $\zeta_\pm$
form overlapping, singly quantized vortex lines with opposite
circulation. The vortex lines are oriented along the $z$ axis, normal
to a dark-soliton plane in the component $\zeta_0$. The structure of
Eq.~(\ref{eq:polar_monopole}) is the analogue of the t'Hooft-Polyakov
monopole.

\subsection*{Construction of prototype interface spinors}

We can construct an approximation of a desired defect configuration that crosses the FM-polar
boundary by starting from a given FM (polar) defect state and inserting
dark-soliton planes in some of the spinor wave function components at the interface. For suitable
choices of spinor parameters, the resulting configuration will then be sufficiently close to
 a local energetic minimum so that it quickly relaxes to the targeted defect structure. We will now
explicitly construct the prototype spinors considered in the main text.

{\em Singular vortex in both phases:} To illustrate the method
outlined above, we create a state with a singly quantized vortex on both
sides of the interface [Fig.~1(a)] by starting from the FM singular vortex,
Eq.~(\ref{eq:fmsingular}), and inserting a soliton plane in $\zeta_-$
\begin{equation}
  \label{eq:1qs}
  \zeta^{\rm 1\leftrightarrow s} =
  \frac{e^{i\varphi}}{\sqrt{2}}\threevec{\sqrt{2}\cos^2\frac{\beta}{2}}
                                       {\sin\beta}
                                       {\pm\sqrt{2}\sin^2\frac{\beta}{2}},
\end{equation}
where the negative sign is used on the polar side of the
interface.  Note that for a general $\beta$ the state created above
the interface does not necessarily have $|\inleva{\bvec{\hat{F}}}|=0$,
and therefore may not strictly be a polar state. However, the state
created has the appropriate vortex structure and very quickly relaxes
to the singly quantized polar vortex.  The choice $\beta=\pi/2$ yields
a spinor which exactly changes from FM to polar.  Analogous
considerations apply to other prototype spinors.

{\em Singular to coreless vortex:} Inserting a soliton plane in
$\zeta_-$ of the singular vortex $\zeta^{\rm 1^\prime}$ in
Eq.~(\ref{eq:p012}), results in the interface spinor
\begin{equation}
  \label{eq:s-cl}
  \zeta^{{\rm 1}\leftrightarrow{\rm cl}} =
  \frac{1}{\sqrt{2}}\threevec{-\sin\beta}
                             {\sqrt{2}e^{i\varphi}\cos\beta}
                             {\pm e^{2i\varphi}\sin\beta},
\end{equation}
where the negative sign is used on the FM side of the
boundary. Choosing $\beta=\pi/4$ or
$\beta=3\pi/4$ yields $|\inleva{\bvec{\hat{F}}}|=1$ on the FM
side. The case $\beta=3\pi/4$
corresponds to the coreless vortex $\zeta^{\rm cl}$ [Eq.~\ref{eq:cl}]
with fountain-like spin profile.

{\em Dirac monopole on the interface:} A Dirac monopole may be placed
on the interface by inserting a soliton plane into $\zeta^{\rm D}$ of
Eq.~(\ref{eq:dirac})
\begin{equation}
  \zeta^{\rm 1 \leftrightarrow D} = \frac{1}{\sqrt{2}}
                  \threevec{\sqrt{2}e^{-2i\varphi}\cos^2\frac{\theta}{2}}
                           {e^{-i\varphi}\sin\theta}
                           {\pm\sqrt{2}\sin^2\frac{\theta}{2}}.
\end{equation}
The spinor after the change of sign has a general structure similar to
Eq.~(\ref{eq:p012}). The constructed
spinor therefore approximates a singular vortex in the polar part,
terminating on the Dirac monopole.  Depending on which side of the
monopole is placed on the polar side of the interface, the singular
Dirac string either terminates on the monopole from the FM side, or
becomes the singular polar vortex. The structure is equivalent to
$\zeta^{{\rm 1}\leftrightarrow{\rm cl}}$, and both spinors are
represented by Fig.~1(b).

{\em Polar monopole on the interface:}
We consider a defect structure with overlapping, singly quantized
vortex lines in $\zeta_\pm$, both oriented normal to the interface
and of opposite circulation, together with $\pi$-phase kinks in $\zeta_+$
and $\zeta_0$. This can be parametrized as
\begin{equation}
  \zeta^{\rm sv\leftrightarrow pm} =
  \frac{1}{\sqrt{2}}\threevec{\mp e^{-i\varphi}\sin\theta}
                             {\sqrt{2}\cos\theta}
	                     {e^{i\varphi}\sin\theta},
\end{equation}
using the positive sign on the FM side. The resulting structure on the
polar side
represents a radial hedgehog of Eq.~(\ref{eq:polar_monopole}). The
continuation of this on the
FM side is similar to the singular spin vortex $\zeta^{\rm sv}$ in
Eq.~(\ref{eq:spinvortex}). Hence we have constructed
an approximation to a spin vortex on the FM side that terminates on
the polar monopole on the polar side [Fig.~1(c)].

{\em Terminating vortices:} Vortices can also be made to terminate at
the interface.
In Eq.~(\ref{eq:s-cl}), and in Fig.~1(b) of the main text, a singular
polar vortex
perforates across the interface to a coreless FM vortex when we insert
a $\pi$-phase
kink in $\zeta_-$. The resulting defect can be cut in half
while preserving the coherent interface with a continuous
order-parameter field across the interface.
We achieve this by inserting an additional phase kink in $\zeta_0$
that allows the vortices
on either side of the interface to move apart
\begin{subequations}
\begin{align}
    \zeta^{\rm cut} &=
    \frac{1}{\sqrt{2}}\threevec{-\sin\beta}
                               {\sqrt{2}e^{i\varphi}\cos\beta}
                               {e^{2i\varphi}\sin\beta},
    \quad {\rm for}\,\,\, z>0 \\
    \zeta^{\rm cut} & =
    \frac{-1}{\sqrt{2}}\threevec{\sin\beta}
                               {\sqrt{2}e^{i\varphi}\cos\beta}
	  		       {e^{2i\varphi}\sin\beta}, \quad {\rm for}\,\,\, z<0,
\end{align}
\end{subequations}
where we may choose $\beta=3\pi/4$ as in Eq.~(\ref{eq:s-cl}).
This can yield the configuration in Fig.~1(e), where the singular
polar vortex and the FM coreless vortex are spatially separated and
both terminate on the interface.  Since the vortex lines in the
individual spinor components terminate on the soliton planes, it is
also possible
to consider a state where a vortex exists only on one side of the
interface, for instance,
\begin{subequations}
\begin{align}
    \zeta^{\rm pv}  & =
    \frac{1}{\sqrt{2}}\threevec{-\sin\beta}
                               {\sqrt{2}e^{i\varphi}\cos\beta}
                               {e^{2i\varphi}\sin\beta},
    \quad {\rm for}\,\,\, z>0 \\
    \zeta^{\rm pv} & =
    -\frac{1}{\sqrt{2}}\threevec{\sin\beta}
                               {\sqrt{2}\cos\beta}
			       {\sin\beta}, \quad {\rm for}\,\,\, z<0.
\end{align}
\end{subequations}

\begin{figure}[t]
  \centering
  \includegraphics[scale=1]{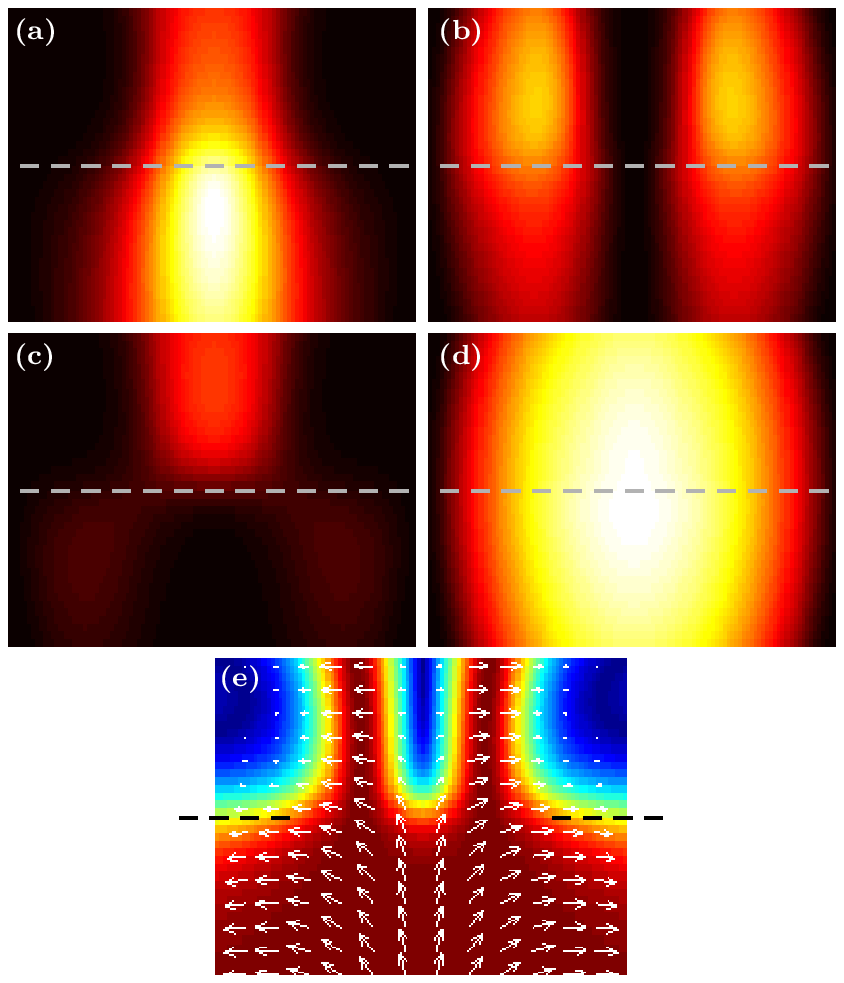}
\caption{Continuity of the order parameter across the interface.
  (a)--(c) Density in the individual spinor components
  $\zeta_+$,$\zeta_0$ and $\zeta_-$, respectively. These do not vanish
  simultaneously on the interface.  The position of the interface,
  where $c_2$ changes sign, is
  indicated by the dashed line.
  (d) The total atom density is continuous across the interface.
  (e) The magnitude (color map; dark red at $\absF=1$) and direction
  (white arrows) of $\inleva{\mathbf{\hat{F}}}$
  vary continuously across the interface (indicated by the dashed
  line; the interface extends across the system, but is indicated only
  on the sides for clarity.)
}
\label{fig:continuity}
\end{figure}

\subsection*{Coherent stable interface}

The stable topological interface between the different ground-state
manifolds of spin-1 BEC 
connects the defects and textures continuously across the interface
boundary. 
Although the 
polar and the FM regions exhibit different broken symmetries and
support different defects 
and textures, the order parameter varies smoothly across the
interface. This is demonstrated 
in Fig.~\ref{fig:continuity} for the energetically stable
configuration of Fig.~2 of the main 
text. 
Note that the total atom density is nonvanishing at the
interface and that both the density 
and the spin varies smoothly across the interface (the position where
$c_2$ changes sign is indicated by a dashed line). The size of the
interface region of the relaxed configuration is given by $\xi_F$ 
representing the healing length of the spin.



%

\end{document}